\documentstyle[12pt]{article}
\begin{document}
\vspace{1.5cm}
\begin{center}
{\large\bf  {QCD STRINGS AS CONSTRAINED \\
\vspace{0.3cm}
     GRASSMANNIAN SIGMA MODEL}} \\
\end{center}
\vspace{3.5cm}

\begin{center}
{\Large {K.S.Viswanathan\footnote{e-mail address: kviswana@sfu.ca}
and R.Parthasarathy\footnote{Permanent address:
The Institute of Mathematical Sciences, Madras 600113,India. \\
e-mail address: sarathy@imsc.ernet.in} \\
Department of Physics \\
Simon Fraser University \\
Burnaby B.C., Canada V5A 1S6.}} \\
\end{center}

\newpage

\vspace{1.5cm}
\noindent{\it Abstract}

\vspace{1.0cm}
We present calculations for the effective action of string world sheet in
$R^3$ and $R^4$ utilizing its correspondence with the constrained
Grassmannian $\sigma$-model. Minimal surfaces describe the dynamics of
open strings while harmonic surfaces describe that of closed strings. The
one-loop effective action for these are calculated with instanton and
anti-instanton background, representing N-string interactions at the tree
level. The effective action is found to be the partition function of a
classical modified Coulomb gas in the confining phase, with a dynamically
generated mass gap.

\newpage

\vspace{0.5cm}

\noindent{\bf I.$\ \ $INTRODUCTION}

\vspace{0.3cm}

Two dimensional non-linear sigma models share many features with \\
4-d Yang-Mills theories. They both are scale-invariant,
asymptotically free and possess multi-instanton solutions [1]. In spite of
these similarities, not much has been done to explore if there is a
deeper relation between these theories. Over the years string models [2]
have been proposed to describe QCD flux tubes, deemed responsible for
quark confinement. It has been widely recognized that QCD strings should
take into account the extrinsic geometry of the string world sheet [3].
General properties of strings with extrinsic curvature action $ S\ =\
\frac{2}{\alpha_0} \int \sqrt{g}{\mid H \mid }^2 d^2\xi $ have been
analyzed [3]. For example, it was shown that this term is asymptotically
free [3,4]. However, so far it has not been established if these rigid
string theories are appropriate to model QCD flux tubes. It is not known
if, for example, they are confining theories.

\vspace{0.5cm}

The present authors have in a series of publications [4-7] studied the
extrinsic geometry of string world sheet immersed in  background
n-dimensional space from the point of view of Grassmannian $\sigma
$-models. The set of all tangent planes to the world sheet of strings
immersed in $R^n$ and regarded as a 2-dimensional Riemann surface endowed
with the induced metric, is equivalent to the Grassmannian manifold $
G_{2,n} \simeq \frac{SO(n)}{SO(2)\times SO(n-2)}\simeq Q_{n-2}\subset
CP^{n-1}$. Note that $G_{2.n}$ can be realized as a quadric $Q_{n-2}$ in
$CP^{n-1}$. It is this representation that we use throughout our work.
However, it is not an ordinary $\sigma$-model, since not every field
in $G_{2,n}$ forms tangent plane to the world sheet. This forces the
$G_{2,n}$ fields to satisfy $(n-2)$ integrability conditions which have
been derived explicitly in [4,6,8,9] by the use of Gauss mapping [8,9].
The Gauss map is a mapping of the tangent planes to the world sheet
$X^{\mu}(z,\bar{z})$ into the Grassmannian $G_{2,n}$ realized as a
quadric in $CP^{n-1}$. There is one third order differential constraint
on the $G_{2,n}$ fields and $(n-3)$ algebraic constraints on the derivative
of the Gauss map. Note that the integrability conditions on the $G_{2,n}$
fields allow us to study the string model in terms of a constrained
$G_{2,n}\ \sigma$-model. These constraints ensure that the $\sigma$-model
does in fact represent a string world sheet. We are stressing this point,
because most other authors incorporate the constraint by requiring that
the normals $N^{\mu}$ ($\sigma$-model fields) to the surface satisfy the
condition ${\partial}_zX^{\mu}\cdot N^{\mu}=0$, thereby making it very
difficult to implement this constraint without dealing with $X^{\mu}$
coordinates. Both the Nambu-Goto (NG) action and the extrinsic
curvature action can be written in terms of their images in $G_{2,n}$
through Gauss maps and the integrability conditions can be implemented by
Lagrange multipliers. Thus the problem of string dynamics visualized as
the dynamics of the world sheet immersed in background $R^n$, can be
transformed into, at least at the classical level, that of a constrained
Grassmannian $\sigma$-model. To complete the picture, the immersion
coordinates $X^{\mu}(z,\bar{z})$ can be reconstructed from the constrained
$G_{2,n}\ \sigma$-model fields [4].

\vspace{0.5cm}

{}From the above discussion, the advantages of studying QCD strings as
a constrained $\sigma$-model should be clear. The extrinsic curvature
action, which usually leads to fourth derivative theory in $X^{\mu}$,
becomes $\sigma$-model action in terms of $G_{2,n}$ fields; the higher
derivatives arising only through the differential integrability
condition. In quantizing this theory, we need to know the proper measure
to use for functional integral over $G_{2,n}$ fields. We recall that
already in the theory of QCD flux tubes, the standard string quantization
cannot be correct [10]. Thus the measure for $X^{\mu}$ integration is not
completely known. Because of this uncertainty, we take the view point
that we can describe QCD flux tubes by constrained $G_{2,4}$
$\sigma$-model ( underlying field theory ) with the usual sigma model
measure. We shall show in this article that the resulting
theory has features similar to the unconstrained $G_{2,4}$
theory, thereby establishing through string description the connection
between QCD and 2-dimensional $\sigma$-models. It will further be seen in
this article that a major advantage of formulating the QCD flux tubes
through Gauss mapping is that it allows one to do path integrals over a
select class of surfaces having prescribed extrinsic geometric properties.

\vspace{0.5cm}

We consider below strings in background $R^3$ and $R^4$ only. The
corresponding Grassmannian manifolds are $G_{2,3}\simeq CP^1$ and
$G_{2,4}\simeq CP^1 \times CP^1$ respectively. Two classes of surfaces,
minimal and harmonic, are considered. Minimal surfaces are non-compact
and have zero scalar mean curvature $h$ [8,9]. They correspond to minimum
action solution to the NG action (area term) i.e. $\Box X^{\mu}\ =\ 0
\Leftrightarrow \ h\ =\ 0$. Minimal surfaces describe the dynamics of
open strings. In the $\sigma$ - model language, minimal surfaces are
described by instantons. In the instanton configuration the $\sigma$ -
model fields are holomorphic. An
$N$-instanton solution that is meromorphic arises as the Gauss map of a
world sheet with $2N$ punctures. This describes $N$-open string
interactions at the tree level (genus zero). Harmonic surfaces correspond
to solutions to the equations of motion of the image of the extrinsic
curvature action in the Grassmannian $G_{2,n}$ [8,9]. However they do not
generally minimize the extrinsic curvature action when expressed  in
terms of $X^{\mu}$ as $\int \sqrt{g} (\Box X^{\mu})^2 d^2\xi $ with
$g_{\alpha \beta}$ as the induced metric. The equation of motion
following from this action is,
\begin{eqnarray}
\sqrt{g}{\Box}^2X^{\mu} + \sqrt{g} \Box X^{\mu}(\Box X^{\nu})^2
+ 2({\partial}_{\bar{z}}X^{\mu})(\Box X^{\nu})({\partial}_z\Box
X^{\nu}) + \nonumber \\
 2({\partial}_zX^{\mu})(\Box X^{\nu})({\partial}_{\bar{z}}\Box
X^{\nu})\ &=&\ 0.
\end{eqnarray}
For immersion in $R^3$ there is only one normal to the surface
defined through $\Box X^{\mu} \ =\ h N^{\mu}$.
The above equation can be
easily generalized to immersion in $R^4$, by using $\Box X^{\mu}\ =\
h_1N_1^{\mu} + h_2N_2^{\mu}$ where $h_1,h_2$ are the projections of
$H^{\mu}$ on to the two normals [4]. Returning to immersion in $R^3$,
for constant $h$ surfaces (1) reads as $\Box
N^{\mu} + h^2 N^{\mu}\ =\ 0$. Expressing $N^{\mu}$ in terms of $G_{2,3}$
fields [4], we find that the this is satisfied only when the $G_{2,3}$
fields are anti-holomorphic. Similar conclusion is reached for immersion
in $R^4$. So, for harmonic surfaces the choice of Grassmannian fields as
antiholomorphic minimizes the extrinsic curvature action whether it is
written in terms of $X^{\mu}$ or the Grassmannian fields. The Gaussian
curvature for these surfaces is a constant and the principal curvatures
are same, thus the world sheet topologically corresponds to a 2-sphere.
Harmonic surfaces describe the dynamics of closed strings. In the
language of the $\sigma$ - model these surfaces correspond to
anti-instantons i.e. $\sigma$ - model fields that are anti-holomorphic.
An N-anti-instanton solution arises as the Gauss map of a world sheet
with 2N punctures and describes N-closed string interactions at the tree
level.

\vspace{0.5cm}

We compute the quantum fluctuations following Fateev, Frolov and \\
Schwarz [11] around instantons (minimal surfaces) and
anti-instantons (harmonic surfaces) in $R^3$ in section II and in
$R^4$ in section III. The resulting effective action is found to be
the partition function for a modified two dimensional classical Coulomb
gas (MCGS) in the plasma phase for immersion of both minimal and harmonic
surfaces in $R^3$ while in $R^4$, it is a MCGS for immersion of minimal
surfaces and a MCGS with interaction between the two $CP^1$-
anti-instantons for immersion of harmonic surfaces; both being in the
confining plasma phase. This means that there is a mass gap for both open
and closed QCD strings which is dynamically generated.

\vspace{1.0cm}

\noindent{\bf II.$\ \ $ QUANTUM FLUCTUATIONS - IMMERSION IN $R^3$}

\vspace{0.5cm}

The Gauss map of a 2-d Riemann surface conformally immersed in $R^3$ has
been considered in detail in Ref.4,8 and 9. By conformal immersion it is
meant that the induced metric is in the conformal gauge
($g_{zz}=g_{\bar{z}\bar{z}}=0;g_{z\bar{z}}\neq 0$), where $z={\xi}_1 +
i{\xi}_2$ with $({\xi}_1,{\xi}_2)$ as coordinates on the world sheet. The
Gauss map is described by,
\begin{eqnarray}
{\partial}_zX^{\mu} & = & \psi \{1-f^2,\
i(1+f^2),\ 2f\},
\end{eqnarray}
where $f$ is the $CP^1$ field and the complex function
$\psi$ is determined by the extrinsic geometry and $f$ [4]. The
NG and extrinsic curvature actions in terms of $G_{2,3}\simeq CP^1
\ \sigma$ -model field $f$ [4] are,
\begin{eqnarray}
S &=& \sigma \int \frac{1}{h^2(z,\bar{z})} \frac{{\mid
f_{\bar{z}}\mid}^2}{(1+{\mid f \mid}^2)^2} \frac{i}{2}dz\wedge d\bar{z}
\nonumber \\
&+& \frac{2}{{\alpha}_0}\int
\frac{{\mid f_{\bar{z}}\mid}^2}{(1+{\mid f\mid}^2)^2} \frac{i}{2}dz\wedge
d\bar{z},
\end{eqnarray}
where $\sigma$ is the string tension and ${\alpha}_0$ is a dimensionless
coupling whose renormalized expression is asymptotically free [4]. The
integrability condition on $f$ is,
\begin{eqnarray}
Im {\left( \frac{f_{z\bar{z}}}{f_{\bar{z}}} -
\frac{2\bar{f}f_z}{1+{\mid f\mid}^2}\right)}_{\bar{z}} &=& 0,
\end{eqnarray}
whenever $f_{\bar{z}}\neq 0$. The notation used throughtout in this paper
is: $f_{\bar{z}}\equiv {\partial}_{\bar{z}}f$. For minimal surfaces,
there is no integrability condition on $f$ and for harmonic surfaces, $f$
satisfies a stronger integrability condition given in (7).

\vspace{1.0cm}

\noindent{\it II.1 $\ \ $ MINIMAL SURFACES IN $R^3$}

\vspace{0.5cm}

As pointed out in the introduction, minimal surfaces in $R^3$ represent the
world sheet of open strings in background 3-d Euclidean space.
For these surfaces the area is minimum and $h\ =\ 0$. The first term in (3)
is the area term $\int \sqrt{g} d^2\xi$. The classical action for the
underlying $G_{2,3}\ \sigma$-model is the second term (3) and classically
there is no integrability on the Gauss map. We consider as a classical
background for the $CP^1$-field, the instanton configuration [11],
\begin{eqnarray}
f(z) &=& c\  \frac{ \prod^N_{i=1} (z-a_i)}{ \prod^N_{i=1}(z-b_i)},
\end{eqnarray}
where $\{c,a_i,b_i\}$ are the (complex) instanton parameters. The quantum
fluctuations $\nu (z,\bar{z})$ around (5) are defined by,
\begin{eqnarray}
f(z)\rightarrow f(z) + \nu (z,\bar{z}).
\end{eqnarray}
The fluctuated field in (6) also arises as the Gauss map of a surface
obtained from the given (classical) minimal surface.
We need the integrability condition on $\nu(z,\bar{z})$. Since
classically there is no integrability condition for minimal surfaces, the
Lagrange multiplier field to implement the condition on $\nu$ must be
quantum. We restrict the fluctuation (6) to correspond to
harmonic surfaces with small but {\it constant $h$}. A Gauss map is said to
be harmonic if it satisfies the Euler-Lagrange equation,
\begin{eqnarray}
f_{z\bar{z}} - \frac{2\bar{f}f_{z}f_{\bar{z}}}{1+{\mid f\mid}^2} &=& 0,
\end{eqnarray}
which is also the equation of motion of the extrinsic curvature action.
It can be shown [4] that if (7) is satisfied then $h$ is constant.
The Gauss map thus satisfies the above stronger integrability condition
whose linearized version with (5) as the classical background is,
\begin{eqnarray}
{\nu}_{z\bar{z}} - \frac{2\bar{f} f_z {\nu}_{\bar{z}}}{1+{\mid f\mid}^2}
&=& 0.
\end{eqnarray}
This is the integrability condition on $\nu$ to be implemented by quantum
multiplier $\lambda$. So the classical minimal surface ($h$ = 0) is
fluctuated to reprsent a (quantum) harmonic surface with small but
constant $h$. Expanding the classical action of the underlying
$G_{2,3}\ \sigma$-model using (5) and implementing the condition (8), we
find,
\begin{eqnarray}
S &=& \frac{2\pi}{{\alpha}_0}N + \frac{4}{{\alpha}_0} \int
\bar{\tilde{\nu}} {\bigtriangleup}_f \tilde{\nu} \sqrt{\cal{G}} d^2z
\nonumber \\
&+& \int \lambda \left( {\nu}_{z\bar{z}} -
\frac{2\bar{f}f_z{\nu}_{\bar{z}}}{1+{\mid f\mid}^2}\right) d^2z + h.c.,
\end{eqnarray}
where,
\begin{eqnarray}
{\bigtriangleup}_f &=& - \frac{1}{\sqrt{\cal{G}}}\rho {\partial}_z
{\rho}^{-2}{\partial}_{\bar{z}} \rho, \nonumber \\
\tilde{\nu} &=& \frac{2\nu}{\rho} \prod^{N}_{i=1} (z-b_i)^2, \nonumber \\
\rho &=& {\rho}_{0} \prod^{N}_{i=1} {\mid z-b_i\mid}^2, \nonumber \\
{\rho}_0 &=& 1+{\mid f\mid}^2,
\end{eqnarray}
where ${\cal{G}}_{\alpha\beta}$ is a metric introduced to avoid infra-red
divergences, which will eventually be taken as ${\delta}_{\alpha\beta}$.
Thus the effective action of open strings in background $R^3$ is given by
(9). The third term in (9) can be simplified to,
\begin{eqnarray}
\int \bar{\tilde{\lambda}} {\bigtriangleup}_f \tilde{\nu} d^2z &+& h.c.,
\end{eqnarray}
where,
\begin{eqnarray}
\tilde{\lambda} &=& \frac{\rho {\lambda}}2{\prod^{N}_{i=1}(z-b_i)^2}.
\nonumber
\end{eqnarray}
The partition function is obtained by the functional integral of the
exponential of $S$ in (9) over $\tilde{\nu},\bar{\tilde{\nu}},{\lambda},
\bar{\lambda}$ with the instanton measure arising from the change of the
sigma model measure $[df][d\bar{f}]$ to a measure in the instanton manifold.
By the standard shift of the integration variables, the functional
integration gives the partition function as,
\begin{eqnarray}
Z &=& \sum_{N=0}^{\infty} (N!)^{-2}
 exp(-\frac{2\pi N}{{\alpha}_0}) \int d{\mu}_0 det
(\frac{4}{{\alpha}_0}{\bigtriangleup}_f)^{-1},
\end{eqnarray}
where $d{\mu}_0$ is the instanton measure [11] and a sum over instantons
of all winding numbers is introduced. The effect of the
integrability condition is to produce additional $(det
{\bigtriangleup}_f)^{-\frac{1}{2}}$ from (11). The determinant
${\bigtriangleup}_f$ and the instanton measure $d{\mu}_0$ have been
evaluated and as this procedure is given in detail in [11], we give only
the results here,
\begin{eqnarray}
Z&=&\sum_{N=0}^{\infty} (N!)^{-2}[(\frac{4}{{\alpha}_0})^2
exp(-\frac{\pi}{{\alpha}_0}) exp(2log\Lambda)]^{2N} \nonumber \\
& &\int \frac{d^2c}{(1+{\mid c\mid}^2)^2} \prod_{j=1}^{N} d^2a_j d^2b_j
\frac{{\mid c\mid}^{-4N}}{(1+{\mid c\mid}^2)^2} (det M)^{-1} \nonumber \\
& &exp[
\sum_{i<j}log{\mid a_i-a_j\mid}^2+\sum_{i<j}log{\mid b_i-b_j\mid}^2
-3\sum_{i,j}log{\mid a_i-b_j\mid}^2].
\end{eqnarray}
This is the partition function for open strings in background $R^3$. In
(13), $\Lambda$ is the effect of ultra-violet cutoff in regularizing the
determinant of ${\bigtriangleup}_f$. This is removed by renormalizing the
coupling constant ${\alpha}_0$ as,
\begin{eqnarray}
{\alpha}_R(\mu) &=& \frac{{\alpha}_0}{1 - 2 (\frac{{\alpha}_0}{\pi})
log{\frac{\Lambda}{\mu}}},
\end{eqnarray}
where $\mu$ is the renormalization point. For unconstrained $CP^1$ model
[11], the factor in front of $(\frac{{\alpha}_0}{\pi})$ is unity. The
role of the integrability condition is thus to change the above factor,
from $(d-2)$ to $(d-1)$, for immersion in $d$ -
dimensional space. This is in agreement with our earlier calculations
[4]. (13) reprsents the classical grand partition function of a two
dimensional modified Coulomb gas system (MCGS), modified in the sense
that the potential energy between oppositely charged particles is
threefold stronger than the repulsive energy between like charges. In the
unconstrained $CP^1$ model, the corresponding partition function is
precisely that of a 2-d Coulomb gas. The modification here is due to the
integrability condition. Furthermore, there is an extra factor of $det M$
where, $M_{ij}\ =\ \int {\bar{z}}^i z^j {\rho}^{-2} d^2z$. We have
checked that the (finite) contribution of $det M$ is independent of the
instanton parameters. Absorbing this and other factors involving $c$, and
denoting them by $\kappa$, we rewrite (13) as,
\begin{eqnarray}
Z &=& \sum_N \int \frac{{\kappa}^N}{(N!)^2}
exp(-E_N(a,b))\prod^{N}_{j=1} d^2a_j d^2b_j,
\end{eqnarray}
where,
\begin{eqnarray}
E_N(a,b)&=& -\sum_{i<j}log{\mid a_i-a_j\mid}^2 - \sum_{i<j}log{\mid
b_i-b_j\mid}^2 \nonumber \\
&+& 3\sum_{i,j}log{\mid a_i-b_j\mid}^2.
\end{eqnarray}
In the case of the $CP^1$ model, it is known that the CGS at the inverse
temperature $\beta\ =\ 1$ is in the plasma phase and has a mass gap. The
critical
temperature for transition to molecular phase has been determined  using
an iterated mean field approximation by Kosterlitz and Thouless [12]. It
is given as solution to
 $q^2 {\beta}_c\ \simeq \ 2\ +\ 2.6\pi exp(-{\mu}'{\beta}_c)$, where
$q^2$ is the strength of the attractive interaction. For
${\mu}'=0$, $\beta_c \simeq 10 \ $, and so the CGS is in the
plasma phase. The MCGS in (16) is again at $\beta \ =\ 1 $, but
with the strength of the attractive interaction $q^2\ =\ 3$. The critical
temperature, is not known for MCGS. However, we can argue
that the effect of the stronger attractive interaction would be to
increase the critical temperature. This qualitative feature is in
agreement with the estimate of $\beta_c$ from the
above expression which gives $\beta_c \simeq\ 3$ for $q^2\ =\ 3$
for zero chemical potential. This suggests that the MCGS in (16) is
in the plasma phase with a mass gap.

\vspace{0.5cm}

\noindent{\it II.2. HARMONIC SURFACES IN $R^3$}

\vspace{0.5cm}

The Gauss map of surfaces immersed in $R^3$ is said to be harmonic if $f$
satisfies (7). For harmonic Gauss maps [8,9], the mean
curvature scalar $h$ is constant and the given surface is compact.
Harmonic maps thus describe closed string dynamics [2]. The integrability
condition for harmonic Gauss maps is also the equation of motion (7). In
order to minimize the extrinsic curvature action in terms of $X^{\mu}$
(see (1)), $f$ in (7) should be anti-holomorphic. This result
is also the content of Chern's theorem [14]. In this case, as mentioned
in the introduction, the surface is actually a Riemann sphere and thus we
are considering closed string dynamics at the tree level. As a solution
to harmonic Gauss map (7), we consider the anti-instanton configuration,
\begin{eqnarray}
f(\bar{z}) &=& \bar{c}\  \frac{\prod^{N}_{i=1} (\bar{z}
-{\bar{a}}_i)}{\prod^{N}_{i=1}(\bar{z} -{\bar{b}}_i)},
\end{eqnarray}
where $\{\bar{c},{\bar{a}}_i,{\bar{b}}_i\}$ are the anti-instanton
parameters. The points $({\bar{a}}_i,{\bar{b}}_i)$ at which the Gauss map
has zeroes and poles represent punctures on the sphere and thus we can
think of inclusion of such configurations as representing N-string
interactions at the tree level. Classically we have $h$ constant and so
redefining $\frac{\sigma}{h^2} + \frac{2}{{\alpha}_0}$ as
$\frac{2}{{\alpha}_0}$, the classical action is,
\begin{eqnarray}
S &=& \frac{2}{{\alpha}_0} \int \frac{ {\mid f_{\bar{z}}\mid}^2}{(1+{\mid
f\mid}^2)^2} \frac{i}{2}dz\wedge d\bar{z} \nonumber \\
&+& \int \lambda \{ f_{z\bar{z}} - \frac{2\bar{f}f_zf_{\bar{z}}}{1+{\mid
f\mid}^2}\} \frac{i}{2}dz\wedge d\bar{z} + h.c.,
\end{eqnarray}
where the integrability condition (7) for harmonic surfaces is implemented
in (18) by a Lagrange multiplier field $\lambda$. For the classical
background (17), the integrability condition (7) is identically satisfied.
The equations of motion of the total classical action contain in addition to
(7) for $f$, a homogeneous equation for $\lambda$. The trivial solution
${\lambda}_{cl}\ =\ 0$ is chosen as this choice is reasonable, for, at
the classical level, the equation of motion for $f$ already ensures the
integrability condition and thus no multiplier field would be needed. The
quantum fluctuation $\nu$,
\begin{eqnarray}
f(\bar{z}) &\rightarrow f(\bar{z}) + \nu(z,\bar{z}),
\end{eqnarray}
describes a surface which is Gauss mapped into $G_{2,3}$. To ensure that
$f(z,\bar{z})$ in (19) also arises as a Gauss map of a surface, we
implement the constraint in (18) in its linearized version
(linear in $\nu$) with $\lambda$ as quantum. Thus the fluctuated surface
is represented by harmonic Gauss map and so its scalar mean curvature $h$
is also constant. Then expanding (18) using (17), we obtain,
\begin{eqnarray}
S &=& \frac{2\pi}{{\alpha}_0} N + \frac{4}{{\alpha}_0}\int
\bar{\tilde{\nu}} {\bigtriangleup}_{f} \tilde{\nu}\sqrt{\cal{G}} d^2z
\nonumber \\
&+& \int \bar{\lambda}\{ {\nu}_{z\bar{z}} -
\frac{2\bar{f}f_{\bar{z}}{\nu}_z}{1+{\mid f\mid}^2}\} d^2z + h.c.,
\end{eqnarray}
where $\tilde{\nu}$ and ${\bigtriangleup}_f$ are same quantities that
appear in (10) with
$z\rightarrow \bar{z},\ b\rightarrow \bar{b}$, and $\frac{i}{2}dz\wedge
d\bar{z}$ is denoted by $d^2z$. The evaluation of
the partition function proceeds exactly the same way as in II.1. The
system is represented by a modified Coulomb gas
in the plasma phase.

\vspace{0.5cm}

In both the cases, the theory of world sheet immersed in $R^3$ is a
modified CGS at an inverse temperature $\beta\ =\ 1$ in the plasma phase
with a dynamically generated mass gap
$m\ =\ \mu exp(-\pi/{\alpha}_R(\mu))$.

\vspace{1.0cm}

\noindent{\bf III.$\ \ $ QUANTUM FLUCTUATIONS-IMMERSION IN $R^4$ \\
                  QCD-STRINGS}

\vspace{1.0cm}

The Grassmannian $\sigma$-model approach to the string dynamics
presented in section II, is extended to string world sheet immersed
in $R^4$. We have two normals $N^{\mu}_i\ (\mu = 1,2,3,4;
i=1,2)$ at each point on the surface and so there are two extrinsic
curvature tensors $H^{\mu}_{\alpha\beta i}$ for the surface. The scalar
mean curvature $h\ =\ \sqrt(h^2_1+h^2_2)$ and the detailed expressions
for $h_1,h_2$ and the extrinsic curvature action in terms of Gauss map
are derived in Ref.4. In the case of $R^4$, $Q_2$ which is equivalent to
$CP^1\ \times\ CP^1$ is parmeterized by $f_1$ and $f_2$ and
the Gauss map is given by,
\begin{eqnarray}
{\partial}_zX^{\mu} &=& \psi [1+f_1f_2,i(1-f_1f_2),f_1-f_2,-i(f_1+f_2)],
\end{eqnarray}
where the complex function $\psi$ is determined by the extrinsic geometry
and $f_1$ and $f_2$ [4]. The two integrability conditions are,
\begin{eqnarray}
Im \left( \sum_{i=1}^2
\frac{f_{iz\bar{z}}}{f_{i\bar{z}}}-\frac{2\bar{f}_if_{iz}}{1+{\mid
f_i\mid}^2} \right)_{\bar{z}} &=& 0,
\end{eqnarray}
and,
\begin{eqnarray}
\mid F_1 \mid &=& \mid F_2 \mid ,
\end{eqnarray}
where $F_i \ =\ f_{i\bar{z}}/(1+{\mid f\mid}^2)$ and whenever
$f_{i\bar{z}}\ \neq \ 0$. In our considerations [4] the world sheet is
described locally by $X^{\mu}(z,\bar{z})$ and the Gauss map allows us to
express NG and extrinsic curvature actions as $G_{2,4}\  \sigma$-model
action,
\begin{eqnarray}
S&=&\int \left(\frac{\sigma}{h^2(z,\bar{z})} + \frac{2}{\alpha_0}\right)
\sum_{i=1}^{2} \frac{{\mid f_{i\bar{z}}\mid}^2}{(1+{\mid f_i\mid}^2)^2}
d^2z.
\end{eqnarray}
This together with (22) and (23) descibes the dynamics of the string
world sheet in background $R^4$.The scalar mean curvature $h$ is given
by [4],
\begin{eqnarray}
(log h)_z&=&\sum_{i=1}^{2}\left( \frac{f_{iz\bar{z}}}{f_{i\bar{z}}} -
\frac{2\bar{f}_i f_{iz}}{1+{\mid f_i\mid }^2}\right),
\end{eqnarray}
and the Gauss map (21) is said to be harmonic if [8,9],
\begin{eqnarray}
f_{iz\bar{z}} - \frac{2\bar{f}_i f_{iz} f_{i\bar{z}}}{1+{\mid f_i
\mid}^2} &=& 0\ ;\ \ \ \ i=1,2.
\end{eqnarray}
{}From (25) and (26) it follows that when the Gauss map is harmonic, $h$ is
constant. But all surfaces of constant $h$ are not necessarily harmonic.
This observation will be used here.

\vspace{0.5cm}

\noindent{\it III.1 $\ \ $ MINIMAL IMMERSION IN $R^4$}

\vspace{0.5cm}

The Gauss map of minimal surfaces ($h\ =\ 0)$ in $R^4$ has been studied
in detail [8,9,14] and accordingly , if $F_1\ =\ F_2 \ \equiv 0$, then
the Gauss map represents a minimal surface in $R^4$, provided the surface is
non-compact. A solution to $F_1\ =\ F_2\ \equiv 0$ is given by
holomorphic functions $f_1(z)$ and $f_2(z)$. There are no integrability
conditions classically. The holomorphic functions $f_1(z)$ and $f_2(z)$
are chosen as,
\begin{eqnarray}
f_i(z) &=& c_i \
\frac{\prod_{j=1}^{N_i}(z-a_{ij})}{\prod_{j=1}^{N_i}(z-b_{ij})};\ \ \
i=1,2.,
\end{eqnarray}
representing background instantons, with parameters
$\{c_i,a_{ij},b_{ij}\}$ for i = 1,2.

\vspace{0.5cm}

Quantum fluctuations ${\nu}_i(z,\bar{z})$ around the instanton background
are defined through,
\begin{eqnarray}
f_i(z) & \rightarrow f_i(z) + {\nu}_i(z,\bar{z}); \ \ \ i=1,2.
\end{eqnarray}
The fluctuated surface also arises as the Gauss map of a surface obtained
from the given classical minimal surface. So we need
integrability conditions on ${\nu}_i$. Since classically there are no
integrability conditions for minimal surface, the Lagrange multiplier
fields to implement the conditions on ${\nu}_i$ must be quantum. In view
of the similar situation in II.1, we restrict the fluctuations (28) to
reprsent harmonic surfaces (i.e. the fluctuated surface has
constant scalar mean curvature). Thus the fields (28) are
required to satisfy (26) in its linearized
version, linear in ${\nu}_i$. For the background (27), the
linearized version of (26) is,
\begin{eqnarray}
{\nu}_{iz\bar{z}} - \frac{2\bar{f}_i f_{iz} {\nu}_{i\bar{z}}}{1+{\mid
f_i\mid }^2} &=& 0; \ \ \ i=1,2.,
\end{eqnarray}
which are implemented by two multipliers. For surfaces immersed in
$R^4$, we need to examine the algebraic integrability condition (23) as
well, for the fields (28). It can be readily checked that there are.
no linear terms in ${\nu}_i$ arising from the constraint (23) in the
instanton background.

\vspace{0.5cm}

As a classical action, we consider the action for the underlying
$G_{2,4}\ \sigma$-model which can be rewritten as,
\begin{eqnarray}
S &=& \frac{2\pi}{{\alpha}_0} (N_1+N_2) + \frac{4}{{\alpha}_0}\int
\sum_{i=1}^{2} \frac{{\mid f_{iz}\mid}^2}{(1+{\mid f_i\mid}^2)^2} d^2z,
\end{eqnarray}
where $N_1, N_2$ are the winding numbers of the two $CP^1$ instantons.
Expanding (30) using (27) and (28), and implementing (29), the effective
action is found to be,
\begin{eqnarray}
S &=& \frac{2\pi}{{\alpha}_0} (N_1+N_2) - \frac{4}{{\alpha}_0} \int
\left( {\bar{\tilde{\nu}}}_1{\bigtriangleup}_{f_{1}}\tilde{\nu} +
{\bar{\tilde{\nu}}}_2 {\bigtriangleup}_{f_{2}} \tilde{\nu}\right) d^2z
\nonumber \\
&+& \int {\bar{\lambda}}_1\{ {\nu}_{1z\bar{z}} -
\frac{2{\bar{f}}_1f_{1z}{\nu}_{1\bar{z}}}{1+{\mid f_1\mid}^2}\} d^2z +
h.c \nonumber \\
&+& \int {\bar{\lambda}}_2\{ {\nu}_{2z\bar{z}} -
\frac{2{\bar{f}}_2f_{2z}{\nu}_{2\bar{z}}}{1+{\mid f_2\mid}^2}\} d^2z +
h.c.,
\end{eqnarray}
where
\begin{eqnarray}
{\bigtriangleup}_{f_{i}} &=& -\frac{1}{\sqrt{\cal{G}}} {\rho}_i
{\partial}_z {\rho}^{-2}_i {\partial}_{\bar{z}}{\rho}_i, \nonumber \\
{\tilde{\nu}}_i &=& \frac{2{\nu}_i}{{\rho}_i} \prod_{j=1}^{N_i}
(z-b_{ij})^2, \nonumber \\
{\rho}_i &=& {\rho}_{0i} \prod_{j=1}^{N_i}{\mid z-b_{ij}\mid }^2,
\nonumber \\
{\rho}_{0i} &=& 1+ {\mid f_i \mid }^2.
\end{eqnarray}
By comparing (31) with (9), it is seen that a doubling corresponding to
the two $CP^1$ instantons occur. The evaluation of the partition function
then is similar as in section II.1, now with measures for the two $CP^1$
instantons and then the result is,
\begin{eqnarray}
Z&=&\sum_{N_1,N_2}\frac{{\kappa}^{N_1+N_2}}{(N_{1}!)^2(N_{2}!)^2}
exp[-E_{N_{1}}(a_1,b_1)-E_{N_{2}}(a_2,b_2)] \nonumber \\
& &
\prod_{j=1}^{N_1}d^2a_{1j}d^2b_{1j}\prod_{k=1}^{N_2}d^2a_{2k}d^2b_{2k},
\end{eqnarray}
where,
\begin{eqnarray}
E_{N_i}(a_i,b_i)&=& -\sum_{k<j}^{N_1} log{\mid
a_{ik}-a_{ij}\mid}^2 - \sum_{k<j}^{N_1} log{\mid
b_{ik}-b_{ij}\mid}^2 \nonumber \\
& + & 3\sum_{k,j}^{N_1} log{\mid a_{ik}-b_{ij}\mid}^2, \ \ \ i=1,2.
\end{eqnarray}
The partition function (33) thus represents two modified Coulomb gas,
arising from the two $CP^1$ instantons. Following the discussion in
section.II, we conclude that there is a mass gap in the quantum theory
of minimal surfaces in $R^4$.

\vspace{1.0cm}

\noindent{\it III.2 $\ \ $ SURFACES OF CONSTANT MEAN CURVATURE}

\vspace{0.5cm}

We consider 2-d surfaces in $R^4$ described by harmonic Gauss map as
representing closed QCD-strings. For harmonic Gauss map, as can be seen
from (25), the scalar mean curvature $h$ is constant. The two
$CP^1$-fields satisfy the harmonic map equation (26). When $h$ is
constant, the NG and extrinsic curvature actions can be written as,
\begin{eqnarray}
S &=& \left( \frac{\sigma}{h^2} + \frac{2}{{\alpha}_0}\right) \int
\sum_{i=1}^{2} \frac{{\mid f_{i\bar{z}}\mid}^2}{(1+{\mid f_i\mid}^2)^2}
d^2z,
\end{eqnarray}
whose equations of motion are the same as (26). However, as pointed out
in the introduction, the extrinsic curvature action expressed in terms of
$X^{\mu}(z,\bar{z})$ acquires a minimum only when $f_1$ and $f_2$ are
anti-holomorphic. In this case, the surface is a compact 2-sphere
representing the world sheet. The two anti-holomorphic functions are
chosen as the two $CP^1$ -anti-instanton configurations,
\begin{eqnarray}
f_i(\bar{z}) &=& {\bar{c}}_i\  \frac{\prod_{j=1}^{N_i}
(\bar{z}-{\bar{a}}_{ij})}{\prod_{j=1}^{N_i}(\bar{z}-{\bar{b}}_{ij})},\ \
\ i=1,2.,
\end{eqnarray}
where $\{{\bar{c}}_i,{\bar{a}}_{ij},{\bar{b}}_{ij}\}$ for i=1,2 are the
anti-instanton parameters. The above background satisfies the harmonic
map equation (26). The algebraic integrability condition (23) has to be
satisfied at
the classical level, in order for (36) to represent a Gauss map. This
puts conditions on the positions of the two $CP^1$ instantons.

\vspace{0.5cm}

The quantum fluctuations are defined by,
\begin{eqnarray}
f_i(\bar{z}) &\rightarrow f_i(\bar{z}) + {\nu}_i(z,\bar{z}); \ \ \ i=1,2.
\end{eqnarray}
We now examine the integrability conditions on (37). We restrict fluctuations
to reprsent surfaces of constant scalar mean curvature. In this case, the
constraint reads as (see (25)),
\begin{eqnarray}
\sum_{i=1}^{2} \left( \frac{f_{iz\bar{z}}}{f_{i\bar{z}}} -
\frac{2{\bar{f}}_i f_{iz}}{1+{\mid f_i\mid}^2}\right) &=& 0,
\end{eqnarray}
whose linearized version is,
\begin{eqnarray}
\sum_{i=1}^{2} \left( \frac{{\nu}_{iz\bar{z}}}{f_{i\bar{z}}} -
\frac{2{\bar{f}}_i {\nu}_{iz}}{1+{\mid f_i\mid}^2}\right) &=& 0.
\end{eqnarray}
The second integrability condition ${\mid F_1\mid}^2 \ =\ {\mid
F_2\mid}^2$ is likewise expanded to give its linearized version.
Expanding (35), using (36) and (37), and implementing (39) along with the
linearized version of ${\mid F_1\mid}^2\ =\ {\mid F_2\mid}^2$, we obtain,
\begin{eqnarray}
S_{eff} &=& \frac{2\pi}{{\beta}_0}(N_1+N_2) - \frac{4}{{\beta}_0}\int
\sum_{i=1}^{2} \left({\bar{\tilde{\nu}}}_i {\bigtriangleup}_{f_{i}}
{\tilde{\nu}}_i\right) d^2z \nonumber \\
&+& \int \lambda
\sum_{i=1}^{2}\left(\frac{{\nu}_{iz\bar{z}}}{f_{i\bar{z}}} -
\frac{2{\bar{f}}_i{\nu}_{iz}}{1+{\mid f_i\mid}^2}\right) d^2z + h.c,
\nonumber \\
&-& \int {\bar{\nu}}_1({\partial}_z\chi)\frac{f_{1\bar{z}}}{(1+{\mid
f_1\mid}^2)^2} d^2z - h.c, \nonumber \\
&+& \int {\bar{\nu}}_2({\partial}_z\chi)\frac{f_{2\bar{z}}}{(1+{\mid
f_2\mid}^2)^2} d^2z + h.c,
\end{eqnarray}
where $\lambda$ and $\chi$ are the quantum multipliers and
$\frac{2}{{\beta}_0} \ =\ \frac{\sigma}{h^2} + \frac{2}{{\alpha}_0}$. The
terms involving the multiplier fields are rewritten using (36) as,
\begin{eqnarray}
S_{eff} &=& \frac{2\pi}{{\beta}_0} (N_1+N_2) - \frac{4}{{\beta}_0} \int
\left( {\bar{\tilde{\nu}}}_1{\bigtriangleup}_{f_{1}} {\tilde{\nu}}_1 +
{\bar{\tilde{\nu}}}_2{\bigtriangleup}_{f_{2}}{\tilde{\nu}}_2\right) d^2z
\nonumber \\
&+&\int{\bar{\tilde{\lambda}}}_1{\bigtriangleup}_{f_{1}}{\tilde{\nu}}_1 +
h.c +\int
{\bar{\tilde{\lambda}}}_2{\bigtriangleup}_{f_{2}}{\tilde{\nu}}_2 + h.c
\nonumber \\
&-& \int {\bar{\tilde{\nu}}}_1{\chi}_1 - h.c + \int
{\bar{\tilde{\nu}}}_2{\chi}_2 + h.c.,
\end{eqnarray}
where,
\begin{eqnarray}
{\tilde{\lambda}}_i &=& \frac{{\rho}_i\lambda}{2\prod_{j=1}^{N_i}
(z-b_{ij})^2 f_{i\bar{z}}}, \nonumber \\
{\chi}_i &=&
\frac{{\rho}_if_{i\bar{z}}{\chi}'}{2{\rho}_{0i}^2\prod_{j=1}^{N_i}
(\bar{z}-{\bar{b}}_{ij})^2}, \nonumber
\end{eqnarray}
with ${\chi}' = {\partial}_z\chi$. The expressions for $\bigtriangleup,
\tilde{\nu}$ in (39) are the same as in (31) with $z\rightarrow \bar{z}\ ,\
b\rightarrow \bar{b}$. The partition function is obtained by the
functional integral of the exponential of $S_{eff}$ over all the quantum
fields. In order to perform this, the quantum action is rewitten, by
shifting the fields, as,
\begin{eqnarray}
S_q &=& -\frac{4}{{\beta}_0} [ \int
({\bar{\tilde{\nu}}}_1-{\beta}_0{\bar{\tilde{\lambda}}}_1+{\beta}_0
{\bar{\xi}}_1) {\bigtriangleup}_{f_{1}}({\tilde{\nu}}_1-{\beta}_0
{\tilde{\lambda}}_1+{\beta}_0{\xi}_1) d^2z \nonumber \\
&+&\int
({\bar{\tilde{\nu}}}_2-{\beta}_0{\bar{\tilde{\lambda}}}_2+{\beta}_0
{\bar{\xi}}_2){\bigtriangleup}_{f_{2}}({\tilde{\nu}}_2-{\beta}_0
{\tilde{\lambda}}_2+{\beta}_0{\xi}_2) d^2z \nonumber \\
&-&\int ({\beta}_0({\bar{\tilde{\lambda}}}_1-{\bar{\xi}}_1))
{\bigtriangleup}_{f_{1}}({\beta}_0({\tilde{\lambda}}_1-{\xi}_1)) d^2z
\nonumber \\
&-&\int
({\beta}_0({\bar{\tilde{\lambda}}}_2+{\bar{\xi}}_2))
{\bigtriangleup}_{f_{2}}({\beta}_0({\tilde{\lambda}}_2+{\xi}_2)) d^2z],
\end{eqnarray}
where ${\xi}_i\ =\ {\bigtriangleup '}_{f_{i}}^{-1}{\chi}_i$, with the
prime denoting the determinant of non-zero eigenvalues of
${\bigtriangleup}_{f_{i}}$. Note that ${\tilde{\lambda}}_1$ and
${\tilde{\lambda}}_2$ as well as ${\xi}_1$ and ${\xi}_2$ are not
independent. The integration over $\{
{\tilde{\nu}}_1,{\tilde{\nu}}_2,\lambda, \chi '\}$ can be transformed
into their linear combinations in (42). This introduces the Jacobian of
the transformation, which is calculated using (36) and (23) as
$det^{-1}({\bigtriangleup '}_{f_{1}}^{-1}\ +\ {\bigtriangleup
'}_{f_{2}}^{-1})$. Thus the partition function for (41) becomes,
\begin{eqnarray}
Z &=& {\beta}_0^{-2(N_1+N_2)} exp(-\frac{2\pi}{{\beta}_0}) \int
det{\bigtriangleup}_{f_{1}}^{-1}\ det{\bigtriangleup}_{f_{2}}^{-1}
\nonumber \\
& & det ({\bigtriangleup '}_{f_{1}}^{-1}\ +\ {\bigtriangleup
'}_{f_{2}}^{-1})^{-1} \ d{\mu}_{01}\ d{\mu}_{02},
\end{eqnarray}
The role of the integrability conditions is first to
produce additional \\
$(det {\bigtriangleup}_{f_{i}})^{-\frac{1}{2}}$ as in
III.1 and secondly to produce $det ({\bigtriangleup '}_{f_{1}}^{-1} +
{\bigtriangleup '}_{f_{2}}^{-1})^{-1}$ from the Jacobian , with the
classical instanton parameters related by (23).

\vspace{0.5cm}

We now discuss the implications of (43). If the Jacobian is ignored, then
we would have obtained MCGS for the two $CP^1$-instantons with no
interaction between them. In this way the integrability condition (23)
couples the two $CP^1$-instantons at the quantum level as well. $det
{\bigtriangleup}_{f_{1}}^{-1}$ and $det{\bigtriangleup}_{f_{2}}^{-1}$ are
evaluated using the methods of Fateev,Frolov and Schwarz [11]. We infer
the form of
$det^{-1}({\bigtriangleup}_{f_{1}}^{-1}+{\bigtriangleup}_{f_{2}}^{-1})$ by
the following procedure. Let us note that the interaction term is symmetric
in the indices 1 and 2. When
${\bigtriangleup}_{f_{1}}\ =\ {\bigtriangleup}_{f_{2}}$, (43) reduces to
the results for immersion in $R^3$ (as it should be) with one instanton
measure removed. The regularized expression for $log det
{\bigtriangleup}_{f_{1}}^{-1}$ is,
\begin{eqnarray}
-4\sum_{j,k} log{\mid
a_{1j}-b_{1k}\mid}^2 &-& 4 log {\mid c_1\mid}^{2N_1}(1+{\mid c_1\mid}^2),
\nonumber
\end{eqnarray}
and similar expression for $log det
{\bigtriangleup}_{f_{2}}^{-1}$, apart from $det M_i$. These observations
suggest that $log det ({\bigtriangleup}_{f_{1}} +
{\bigtriangleup}_{f_{2}}^{-1})$ should be of the form,
\begin{eqnarray}
2 \sum_{j,k} log{\mid a_{1j}-b_{2k}\mid}^2 &+& 2 \sum_{j,k} log {\mid
a_{2j}-b_{1k}\mid}^2 \nonumber \\
-A \sum_{j,k} log {\mid
a_{1j}-a_{2k}\mid}^2 &-& B \sum_{j,k} log {\mid b_{1j} - b_{2k}\mid }^2,
\end{eqnarray}
apart from $c_{1,2}$ factors. Then when $a_{1j}\ =\ a_{2k}\
;\ b_{1j} \ =\ b_{2k}$ and with one instanton measure removed, we recover
the results for immersion in $R^3$, modulo an infinite constant. In (44) we
have included the last two terms as possible additional interactions among
the two $CP^1$ - anti-instantons with arbitrary coefficients. Then, the
partition function $Z$ can be written as,
\begin{eqnarray}
Z&=&
\sum_{N_1,N_2} \frac{{\kappa}^{N_1+N_2}}{(N_1!)^2(N_2!)^2}\ exp[
-E_{N_1}(a_1,b_1) - E_{N_2}(a_2,b_2) - V_{N_1,N_2}] \nonumber \\
& & \prod_{j=1}^{N_1} d^2a_{1j} d^2b_{1j}\ \prod_{k=1}^{N_2} d^2a_{2k}
d^2b_{2k},
\end{eqnarray}
where,
\begin{eqnarray}
E_{N_1}(a_1,b_1) &=&
-\sum_{i<j}^{N_1} log {\mid a_{1i} - a_{1j}\mid}^2 - \sum_{i<j}^{N_1}
log{\mid b_{1i} - b_{1j}\mid}^2 \nonumber \\
&+& 3\sum_{i,j}^{N_1} log
{\mid a_{1i} - b_{1j}\mid }^2, \nonumber \\
E_{N_2}(a_2,b_2) &=&
-\sum_{i<j}^{N_2} log {\mid a_{2i} - a_{2j}\mid}^2 - \sum_{i<j}^{N_2}
log{\mid b_{2i} - b_{2j}\mid}^2 \nonumber \\
&+& 3\sum_{i,j}^{N_2} log
{\mid a_{2i} - b_{2j}\mid}^2,  \nonumber \\
V_{N_1,N_2} &=&
-2\sum_{i,j}^{N_!,N_2} log {\mid a_{1i}-b_{2j}\mid}^2 -
2\sum_{i,j}^{N_1,N_2}{\mid a_{2i}-b_{1j}\mid}^2 \nonumber \\
&+& A\sum_{i,j}^{N_1,N_2} log{\mid a_{1i}-a_{2j}\mid}^2 +
B\sum_{i,j}^{N_1,N_2}log{\mid b_{1i}-b_{2j}\mid}^2,
\end{eqnarray}
where the $c_1,c_2$ dependent terms and others have been absorbed in
$\kappa$. The partition function (44) thus represents two MCGS with
possible interactions between them.
In the absence of interactions, the system is in the plasma phase, as can
be seen from III.1. The effect of the interactions between the two $CP^1$ -
anti-instantons in (45) is not expected to change the transition
temperature. A detailed study of
the interaction in (45) will shed further light
on the phase transition.

\vspace{0.5cm}

We now comment on the renormalization of the coupling constant
${\beta}_0$ for immersion in $R^4$. Regularization of the determinants
in (43) leads to the following result for ${\beta}_R(\mu)$,
\begin{eqnarray}
{\beta}_R(\mu) &=& \frac{{\beta}_0}{1 - 3
\left(\frac{{\beta}_0}{\pi}\right) log{\frac{\Lambda}{\mu}}}.
\end{eqnarray}
The factor 3 in the denominator in (47) is consistent with our
observation following (14). For surfaces immersed in $R^4$, the
self-intersection number, a topological invariant, in terms of
the Gauss map (21) is,
\begin{eqnarray}
\cal{I} &=& \frac{1}{2}\int \{ {\mid F_1\mid}^2 - {\mid{\hat{F}}_1\mid}^2
- ({\mid F_2\mid}^2 - {\mid{\hat{F}}_2\mid}^2)\} d^2z,
\end{eqnarray}
where ${\hat{F}}_i\ =\ f_{iz}/(1+{\mid f_i\mid}^2)$; i=1,2 and this
vanishes identically for the background configuration studied here. The
self-intersection number plays the role of the $\theta$-term in the QCD
lagrangian [13]. In order to study the $\theta$-term, we need to consider
immersed surfaces of non-zero self-intersection number.

\vspace{1.0cm}

\noindent{\bf{IV.$\ \ $ SUMMARY AND CONCLUSIONS}}

\vspace{0.5cm}

We have presented calculations for the effective action of the string
world sheet in $R^3$ and $R^4$ utilizing its correspondence with the
constrained Grassmannian $\sigma$ - model. Two classes of surfaces, i.e.
minimal ($h\ =\ 0$) and harmonic ($h\ =\ constant\ \neq \ 0$) which
describe respectively, the dynamics of N-string interactions at the tree
level of open and closed strings are studied at the one loop level. The
effective action is found to be the classical partition function of a
modified 2-d Coulomb gas at an inverse temperature $\beta \ =\ 1$ and it
is found that the system is in the confining (plasma) phase with a mass
gap.

\vspace{0.5cm}

We now make qualitative estimates on the mass ratio of the glue-ball to a
rho meson. The mass gap generated in the case of open-string dynamics
(III.1) is $m_{open}\ =\ \mu exp(-\pi/{\alpha}_R(\mu))$ where
${\alpha}_R(\mu)\ =\ {\alpha}_0/(1-2{\alpha}_0/\pi
log{\frac{\Lambda}{\mu}})$. The bare coupling ${\alpha}_0$ is due to the
extrinsic curvature action alone, since for minimal surfaces $h\ =\ 0$.
However, as pointed out in III.1, the quantum fluctuations are restricted
to constant scalar mean curvature surfaces. Thus, the effective
action is governed by a coupling constant $\frac{\sigma}{h_q^2} +
\frac{2}{{{\alpha}_0}^2}$. This allows us to write,
\begin{eqnarray}
m_{open} \simeq \mu exp(-\frac{\sigma}{2h_q^2} - \frac{1}{{\alpha_0}^2} +
\frac{1}{4}log{\frac{\Lambda}{\mu}}),
\end{eqnarray}
where $h_q$ is small but constant scalar mean curvature of the quantum
surface. The first excited state in the case of open QCD-strings
corresponds to a pion. However as we do not have chiral symmetry breaking
mechanism in our approach, we take rho meson to be the first excited state.
In the case of closed strings, the classical surface has a constant
non-zero $h$ (III.2) and so the fluctuated surface effectively has
constant $h\ +\ h_q$ mean curvature. $2/{{\beta}_0}^2$ in (40) will
effectively be replaced by $\frac{\sigma}{(h+h_q)^2} + \frac{2}
{{\alpha_0}^2}$. The mass gap then is,
\begin{eqnarray}
m_{closed}&\simeq & \mu
exp(-\frac{\sigma}{(h+h_q)^2}-\frac{1}{{\alpha_0}^2} +
\frac{1}{4}log\frac{\Lambda}{\mu}),
\end{eqnarray}
where the same renormalization point as in (49) is used. As the lowest
hadronic state of closed QCD-string corresponds to a glue-ball, we deduce
that,
\begin{eqnarray}
\frac{{m}_{glueball}}{m_{rho}}&\sim &
exp(\frac{\sigma}{2}(\frac{1}{h_q^2} - \frac{1}{(h+h_q)^2}))\ >\ 1.
\end{eqnarray}
This estimate seems to agree with the current thinking on the subject.

\vspace{0.5cm}

\noindent{\bf Acknowledgement}

This work has been supported in part by an operating grant from the
National Sciences and Engineering Council of Canada. We thank H.Trottier
for discussions. R.P thanks the Department of Physics for kind
hospitality and facilities offerred.

\vspace{1.0cm}

\noindent{\bf REFERENCES}

\vspace{0.5cm}

\begin{enumerate}

\item A.M.Polyakov, Phys.Lett.{\bf 59B}(1975)79,82.\\
      A.A.Belavin and A.M.Polyakov, JETP Lett.{\bf 22}(1975)245.\\
      A.D'Adda, M.L\"{u}scher and P.Di Vecchia,Nucl.Phys.{\bf 146B}\\
      (1978) 63. \\

\item M.B.Green, J.H.Scwartz and E.Witten,{\bf Superstring Theory}, \\
      Vols.1,2, Cambridge Univ.Press, Cambridge, 1987. \\

\item A.M.Polyakov, Nucl.Phys. {\bf 268B} (1986) 406; \\
      {\bf Gauge Fields and Strings}, (Harwood), NY, 1987. \\
      J.Ambjorn and B.Durhuus, Phys.Lett. {\bf 188B} (1987) 253. \\

\item K.S.Viswanathan, R.Parthasarathy and D.Kay, \\
      Ann.Phys.(NY) {\bf 206} (1991) 237. \\

\item R.Parthasarathy and K.S.Viswanathan, Int.J.Mod.Phys.{\bf A7}  \\
      (1992) 317. \\

\item R.Parthasarathy and K.S.Viswanathan, Int.J.Mod.Phys. {\bf A7} \\
      (1992) 1819. \\

\item K.S.Viswanathan and R.Parthasarathy, Int.J.Mod.Phys. {\bf A7} \\
      (1992) 5995. \\

\item D.A.Hoffman and R.Osserman, J.Diff.Geom. {\bf 18} (1983) 733. \\

\item D.A.Hoffman and R.Osserman, Proc.London.Math.Soc.(3) {\bf 50} \\
      (1985) 21. \\

\item J.Polchinski and A.Strominger, Phys.Rev.Lett. {\bf 67} (1991) \\
       1681. \\

\item V.A.Fateev, I.V.Frolov and A.S.Schwarz, Nucl.Phys. {\bf 154B} \\
      (1979) 1 ; B.Berg and M.L\"{u}scher, Comm.Math.Phys. {\bf 69} \\
      (1979) 57. \\

\item I.M.Kosterlitz and D.I.Thouless, J.Phys. {\bf 6C} (1973) 1181. \\
      I.M.Kosterlitz, J.Phys. {\bf 7C} (1974) 1046. \\

\item See A.M.Polyakov in Ref.3 \\
      and P.O.Mazur and V.P.Nair, Nucl.Phys.{\bf 284B} (1986) 146; \\
       G.D.Robertson, Phys.Lett. {\bf 226B} (1989) 244. \\

\item J.Eells and A.Ratto, {\bf Harmonic Maps and Minimal \\
      Immersions with Symmetries}, Annals of Mathematical \\
       Studies, Princeton Univ.Press, Princeton. (NJ). 1993. \\
      D.T.Thi and A.T.Fomenko, {\bf Minimal surfaces, Stratified \\
       Multi-Varifolds and the Plateau Problem}, AMS.Vol.84.1991. \\
      R.Osserman, {\bf A Survey of Minimal Surfaces}, Van Nostrand \\
      Reinhold Math.Studies. No.25. 1969.
\end{enumerate}
\end{document}